\documentclass[conference]{IEEEtran}
\IEEEoverridecommandlockouts

\usepackage{color}

\definecolor{AlirezaPurple}{RGB}{150, 0, 250}

\usepackage{cite}
\usepackage{amsmath,amssymb,amsfonts}
\usepackage{algorithmic}
\usepackage{acronym}
\usepackage{graphicx}
\usepackage{textcomp}
\usepackage{xcolor}
\def\BibTeX{{\rm B\kern-.05em{\sc i\kern-.025em b}\kern-.08em
    T\kern-.1667em\lower.7ex\hbox{E}\kern-.125emX}}
\usepackage{geometry}
\acrodef{ad}[AD]{autonomous drive}
\acrodef{alb}[ALB]{absolute lower bound}
\acrodef{adas}[ADAS]{advanced driver assistance system}
\acrodef{aoa}[AOA]{angles-of-arrival}
\acrodef{aod}[AOD]{angles-of-departure}\acrodef{bs}[BS]{base station}
\acrodef{bs}[BS]{base station}
\acrodef{cdf}[CDF]{cumulative distribution function}
\acrodef{crb}[CRB]{Cram\'er–Rao bound}
\acrodef{mcrb}[MCRB]{Mismatched Cram\'er–Rao bound}
\acrodef{dbscan}[DBSCAN]{density-based spatial clustering of applications with noise}
\acrodef{gnss}[GNSS]{global navigation satellite system}
\acrodef{gps}[GPS]{global positioning system}
\acrodef{imu}[IMU]{inertial measurement unit}
\acrodef{isac}[ISAC]{integrated sensing and communication}
\acrodef{ip}[IP]{incidence point}
\acrodef{las}[L\&S]{localization and sensing}
\acrodef{los}[LOS]{line-of-sight}
\acrodef{mae}[MAE]{mean absolute value}
\acrodef{mc}[MC]{mutual coupling}
\acrodef{mcrb}[MCRB]{misspecified Cram\'er–Rao bound}
\acrodef{mcm}[MCM]{mutual coupling matrix}
\acrodef{map}[MAP]{maximum a posteriori}
\acrodef{mle}[MLE]{maximum likelihood estimator}
\acrodef{mpc}[MPC]{multipath component}
\acrodef{nlos}[NLOS]{non-line-of-sight}
\acrodef{nc}[NC]{non-ideal codebook}
\acrodef{ci}[CI]{combined impacts}
\acrodef{ofdm}[OFDM]{orthogonal frequency division multiplexing}
\acrodef{prs}[PRS]{positioning reference signal}
\acrodef{pdf}[PDF]{probability density function}
\acrodef{pss}[PSS]{primary synchronization signal}
\acrodef{rf}[RF]{radio frequency}
\acrodef{ris}[RIS]{reconfigurable intelligent surface}
\acrodef{rev}[REV]{rotating element electric field vector}
\acrodef{rss}[RSS]{received signal strength}
\acrodef{rtk}[RTK]{real-time kinematic}
\acrodef{rtt}[RTT]{round-trip-time}
\acrodef{slam}[SLAM]{simultaneous localization and mapping}
\acrodef{ssb}[SSB]{synchronization signal/physical broadcast channel block}
\acrodef{siso}[SISO]{single-input-single-output}
\acrodef{snr}[SNR]{signal-to-noise ratio}
\acrodef{tdoa}[TDOA]{time-difference-of-arrival}
\acrodef{toa}[TOA]{time-of-arrival}
\acrodef{ue}[UE]{user equipment}
\acrodef{ura}[URA]{uniform rectangular array}
\acrodef{va}[VA]{virtual anchor}

\long\def\comment#1{}

\DeclareMathOperator*{\argmin}{arg\,min}

\newfont{\bbb}{msbm10 scaled 700}

\newfont{\bb}{msbm10 scaled 1100}
\newcommand{\CC}{\mbox{\bb C}}

\newcommand{\av}{{\bf a}}
\newcommand{\bv}{{\bf b}}
\newcommand{\cv}{{\bf c}}
\newcommand{\dv}{{\bf d}}

\newcommand{\mv}{{\bf m}}

\newcommand{\pv}{{\bf p}}

\newcommand{\sv}{{\bf s}}

\newcommand{\wv}{{\bf w}}

\newcommand{\yv}{{\bf y}}

\newcommand{\Am}{{\bf A}}
\newcommand{\Bm}{{\bf B}}
\newcommand{\Cm}{{\bf C}}
\newcommand{\Dm}{{\bf D}}

\newcommand{\Id}{{\bf I}}
\newcommand{\Jm}{{\bf J}}

\newcommand{\Mm}{{\bf M}}

\newcommand{\Qm}{{\bf Q}}

\newcommand{\Wm}{{\bf W}}

\newcommand{\Xm}{{\bf X}}
\newcommand{\Ym}{{\bf Y}}

\newcommand{\etav}{\hbox{\boldmath$\eta$}}

\newcommand{\epsilonv}{\hbox{\boldmath$\epsilon$}}

\newcommand{\muv}{\hbox{\boldmath$\mu$}}

\newcommand{\phiv}{\hbox{\boldmath$\phi$}}

\newcommand{\thetav}{\hbox{$\boldsymbol\theta$}}

\newcommand{\varphiv}{\hbox{\boldmath$\varphi$}}

\newcommand{\Gammam}{\hbox{\boldmath$\Gamma$}}

\renewcommand{\arg}{{\hbox{arg}}}

\begin{document}

\bstctlcite{IEEEexample:BSTcontrol}

\title{RIS Beam Calibration for ISAC Systems: \\ Modeling and Performance Analysis
\thanks{This work was supported, in part by the research grant (VIL59841) from VILLUM FONDEN, the Swedish Research Council (VR grant 2022-03007), the SNS JU project 6G-DISAC under the EU’s Horizon Europe research and innovation programme under Grant Agreement No 101139130.}
}

\author{Mengting Li$^{\dagger,\star}$, Hui Chen$^{\star}$, Sigurd Sandor Petersen$^{\dagger}$, Huiping Huang$^{\star}$, Alireza Pourafzal$^{\star}$, \\ 
Yu Ge$^{\star}$, Ming Shen$^{\dagger}$, Henk Wymeersch$^{\star}$ \\ \newline \vspace{-3mm} \\

$^{\star}$Chalmers University of Technology, Sweden ~ $^{\dagger}$Aalborg University, Denmark 
\vspace{-3mm}

}

\maketitle
\begin{abstract}
High-accuracy localization is a key enabler for integrated sensing and communication (ISAC), playing an essential role in various applications such as autonomous driving. Antenna arrays and reconfigurable intelligent surface (RIS) are incorporated into these systems to achieve high angular resolution, assisting in the localization process. 
However, array and RIS beam patterns in practice often deviate from the idealized models used for algorithm design, leading to significant degradation in positioning accuracy. This mismatch highlights the need for beam calibration to bridge the gap between theoretical models and real-world hardware behavior.
In this paper, we present and analyze three beam models considering several key non-idealities such as \textit{mutual coupling}, \textit{non-ideal codebook}, and \textit{measurement uncertainties}. Based on the models, we then develop calibration algorithms to estimate the model parameters that can be used for future localization tasks.
This work evaluates the effectiveness of the beam models and the calibration algorithms using both theoretical bounds and real-world beam pattern data from an RIS prototype. 
The simulation results show that the model incorporating \textit{combined impacts} can accurately reconstruct measured beam patterns. This highlights the necessity of realistic beam modeling and calibration to achieve high-accuracy localization.
\end{abstract}


\section{Introduction}
Future wireless communication systems are evolving beyond ubiquitous connectivity to emphasize high-accuracy and reliable localization, allowing \ac{isac} capabilities. This shift will enhance conventional location-based services and enable emerging applications such as autonomous driving and healthcare monitoring~\cite{tataria20216g}. The antenna arrays employed by the \ac{bs} provide flexible beamforming capabilities, enabling high angular resolution for angle estimation during the localization process. In addition, the low-cost design of \ac{ris} ~\cite{bjornson2022reconfigurable}, with wave control capabilities, makes it particularly attractive to improve positioning accuracy~\cite{chen2024multi}, reliability, and coverage~\cite{liu2021reconfigurable}, especially in challenging environments where traditional approaches have limited effectiveness.  

A major challenge in achieving high-precision localization arises from the mismatch between practical array or \ac{ris} beams and oversimplified beam models assumed in the localization algorithms~\cite{ghazalian2024calibration}. Typically, the \ac{ris} or array model can be simplified as an antenna array composed of a certain number of isotropic elements. \ac{ris} typically works as a passive array, which can reflect or redirect wireless signals that impinge on its surface, whereas antenna arrays can transmit signals directly. 
The practical \ac{ris} or array beam pattern is affected by several key factors~\cite{tang2019wireless,gradoni2021end}, including \ac{mc} among elements, non-ideal radio frequency (RF) chains (e.g., the phase shift module connected to each element) and non-isotropic element patterns, where the reflection coefficients vary with the angle of departure. Moreover, fabrication errors further contribute to deviations from the idealized model. The model mismatch in the localization systems could have significant impacts on the estimation accuracy \cite{chen2023modeling}. Hence, the use of oversimplified beam models and the resulting inaccuracies in the RIS or array beam pattern may degrade the localization accuracy. Beam calibration, which involves accurately modeling and configuring the realistic characteristics of the deployed \ac{ris} or arrays, is essential for achieving high precision localization.

\begin{figure}[t]
\centering
\centerline{\includegraphics[width=0.8\linewidth]{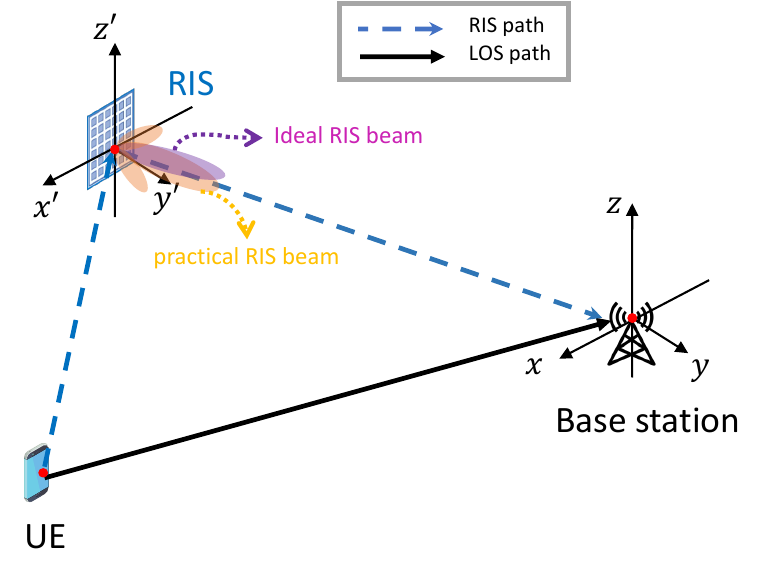}}
\vspace{-0.25cm}
\caption{An illustration of an RIS-aided localization system, where the received signals from both the LOS path and the RIS-reflected path are utilized to determine the position of UE. The ideal \ac{ris} beam represents a simplified beam model used in the localization algorithm, which does not accurately reflect the characteristics of the practical \ac{ris} beam.}
\label{fig_illustration}
\vspace{-0.35cm}
\end{figure}
Array calibration in the state-of-the-art refers to estimating the excitation coefficients and compensating for nonidealities caused by inhomogeneities among RF chains. By assigning dedicated phase shifts to the array elements and recording the received complex signals, the excitation coefficients can be determined by solving the corresponding linear equations \cite{long2017multi}. In \cite{zhang2019improved}, an improved array calibration method was proposed using Hadamard matrix-based phase shifts. In addition, in situ calibration methods that require a minimum number of complex signal measurements for beam-steering arrays are introduced in \cite{wang2021over}. Array calibration can also be achieved using amplitude-only data, and a widely recognized approach is the rotating element electric field vector (REV) method, which measures the power of the received signal while changing the phase of one element over the range of [0$^\circ$, 360$^\circ$] per measurement \cite{takahashi2006theoretical}. Several enhanced versions of REV were proposed to improve the measurement efficiency \cite{takahashi2012novel,long2016fast}. Calibration in \ac{isac} systems could involve characterizing and correcting geometric and beam response inconsistencies in antenna arrays or \ac{ris}. In \cite{zhang2023over}, an over-the-air \ac{ris} calibration method was introduced, using an alternating block descent approach to calibrate the \ac{ris} coefficients. In \cite{zheng2023jrcup}, the uncertainties in \ac{ris} location and orientation of \ac{ris} were considered in the context of multi-user positioning. However, limited research has focused on calibrating \ac{ris} or array beams in these systems.

In this work, we formulate the beam calibration problem in an RIS-aided \ac{isac} system and propose reasonable beam models and corresponding calibration algorithms to improve localization performance. The main contributions are: (i) We present three beam models which consider practical non-idealities such as \ac{mc}, inhomogeneities among \ac{ris} RF branches, and element pattern. (ii) We develop effective calibration algorithms to estimate the unknown parameters in the beam models presented. (iii) We measured the beam patterns of an \ac{ris} prototype and use these data as ground truth to evaluate and validate the proposed beam models and the calibration algorithms through the theoretical lower bounds of the positioning error.

\section{Problem Statement}\label{sec:problem_statement}

\subsection{ISAC System Model}
We consider a three-dimensional (3D) \ac{siso} RIS-aided localization system, as shown in Fig. \ref{fig_illustration}. The base station (BS) is located at the origin $[0,0,0]^\top$ of the global coordinate system, and a \ac{ue} is located at $\mathbf{p}_u = [x_u, y_u,z_u]^\top$. Assume that \ac{ris} consists of ${N} = N_1 \times N_2$ elements centered at $\mathbf{p}_r = [x_r, y_r,z_r]^\top$ under the far-field assumption. RIS elements are placed on the $x'oz'$-plane, with an element spacing of $\frac{\lambda_c}{2}$ along both the $x'$ and $z'$ axes, where $\lambda_c$ denotes the wavelength at the center frequency. For an \ac{ofdm} system consisting of $K$ subcarriers, we assume that $G$ codewords are adopted sequentially for positioning during the coherence time, the received signals $\Ym\in \mathbb{C}^{G\times K}$ can be expressed as \cite{keykhosravi2021siso} %
\begin{align}\label{eq:received_signal}
    \mathbf{Y} &= (\mathbf{H}_{L} + \mathbf{H}_{R}) \odot \mathbf{X} + \Qm \notag \\
    &= ( \underbrace{\alpha_l\mathbf{1}_G\dv^\top(\tau_l)}_{\text{LOS channel}} + \underbrace{\alpha_r\bv(\varphiv_A, \varphiv_D)\dv^\top(\tau_r)}_{\text{RIS channel}} ) \odot \mathbf{X} + \Qm ,
\end{align}
where $\alpha_l$ and $\alpha_r$ are complex channel gains for \ac{los} channel and \ac{ris} channel, respectively and $\tau_l$ and $\tau_r$ are their corresponding delays. Each element in $\Xm$ denotes the transmitted signal symbol and $\Qm$ is the additive white Gaussian noise matrix with each element $q_{g,k}\in\mathcal{CN}(0, \sigma_q^2)$. The beam response vector (for the $G$ codewords) is indicated as $\bv(\varphiv_A, \varphiv_D) = [b_{1}(\varphiv_A, \varphiv_D), b_{2}(\varphiv_A, \varphiv_D) \ldots, b_{G}(\varphiv_A, \varphiv_D)]^\top$ where $\varphiv_A$ and $\varphiv_D$ represent the \ac{aoa} and \ac{aod} at RIS, respectively. Moreover, $\dv(\tau) \in \CC^{K \times 1}$ describes the phase change over different sub-carriers with $d_k(\tau) = e^{j2\pi k \Delta_f \tau}$, determined by delay $\tau$ and subcarrier spacing~ $\Delta_f$.
In most existing studies, the beam model of an \ac{ris} or antenna array in an \ac{isac} system is typically expressed as 
\begin{equation}\label{eq:b_ideal}
    \bv_{\mathrm{ideal}}(\varphiv_A, \varphiv_D) = \Wm^\top(\av(\varphiv_A) \odot \av(\varphiv_D)),
\end{equation}  
where \( \av(\varphiv) \in \CC ^ {{N} \times 1} \) denotes the steering vector, defined as  $[\av(\phi,\theta)]_{{n}} = e^{j \pi (n_1 \sin(\theta)\cos(\phi) + n_2\cos(\theta)}$ (the ${n}$-th RIS element locates at the $n_1$-th column and $n_2$-th row), and $\Wm = [\wv_1, \wv_2, \ldots, \wv_G] \in \CC ^ {{N} \times G}$ is the predefined codebook matrix with each codeword beamforming to a specific direction. 

The core task of this RIS-aided \ac{isac} system is to accurately estimate the position of \ac{ue} based on received signals. However, this simplified beam model may not accurately capture the true characteristics of \ac{ris} or array beam patterns.
\subsection{Objective}
A key prerequisite for achieving high accuracy localization is the development of an accurate and realistic system model. Although the simplified model in (\ref{eq:b_ideal}) offers traceability, it overlooked several important factors, leading to a non-negligible mismatch between the assumed beam model and the actual \ac{ris} beam pattern. This mismatch can degrade localization performance, requiring the development of a more realistic beam model and practical calibration methods. The objective of this work is to explore an effective \ac{ris} beam model considering non-idealities and propose corresponding calibration algorithms aimed at improving the accuracy and reliability of RIS-aided localization.
\section{Beam Pattern Modeling}\label{sec:beam_model}
 While the system is initially formulated in 3D, for simplicity and to maintain consistency with the measurement data presented in Section \ref{sec:validation}-A, the beam pattern modeling in this section focuses on the 2D azimuth plane. Extension to the 3D case is left for future work. It is important to note that the beam modeling discussed here is concerned solely with the intrinsic characteristics of the \ac{ris} or antenna array, independent of any channel effects in the \ac{isac} system. Therefore, the 2D beam models discussed here are functions of the observation angle $\phi$ in the azimuth plane, equivalent to modeling the measured beam patterns in an anechoic chamber. The model in \eqref{eq:b_ideal} can be then updated as 
 \begin{equation}\label{eq:b_ideal_2}
    \bv_{\mathrm{ideal}}(\phi) = \Wm^\top\av(\phi),
\end{equation} 
where the steering vector \( \av(\phi) \in \CC ^ {N \times 1} \) is defined as  $[\av(\phi)]_{n} = e^{j \pi n \cos(\phi)}$ and the codebook matrix $\Wm \in \CC ^ {N \times G}$.
 Compared to the idealized beam model, we account for the effects of \ac{mc}, non-ideal RF components, and individual element patterns in the following discussion. There are various ways \cite{zhang2022phased,chen2018review,wei2018online} 
 to incorporate these effects into the models. Here, we progressively introduce three beam models of increasing complexity to investigate the required model accuracy to improve localization performance.
\subsection{Beam Model with Mutual Coupling Matrix}
\ac{mc} among \ac{ris} elements arises from induced currents or voltages in adjacent elements \cite{li2020split}. As a result, it is more pronounced between closely spaced neighboring elements and can be neglected when the elements are sufficiently far apart. A popular way to model the \ac{mc} effects is to introduce a \ac{mcm} \cite{wei2018online}. While the classical \ac{mcm} typically remains unchanged across transmissions \cite{chen2018off}, here we employ distinct \acp{mcm} for different codewords. This approach enhances model accuracy, as the \ac{mc} within an array can vary with changes in excitation coefficients~\cite{chen2018review}. The beam model with the \ac{mcm} can be formulated as
\begin{align}\label{eq:b_1}
\bv_{\text{MCM}}(\phi) &= [b_1(\phi), b_2(\phi), \cdots, b_G(\phi)]^\top, \notag \\
b_g(\phi) &= (\Cm_g \wv_g)^\top \av(\phi),
\end{align}
where the \ac{mcm} for the $g$-th codeword $\Cm_g \in \mathbb{C}^{N \times N}$ is determined by several \ac{mc} coefficients as
\begin{equation}
    \Cm_g = \mathrm{Toeplitz}([1, c_{g,1}, \ldots, c_{g,m}, 0, \ldots, 0]).
\end{equation}
Here, $c_{g,i}$ is the \ac{mc} coefficient accounting for the \ac{mc} effects between elements with $i$ spacing units for the $g$-th codeword. 
\subsection{Beam Model with Non-ideal Codebook}
The ideal \ac{ris} coefficients are phase adjustments applied to the reflected signals at each \ac{ris} element, determined by a predefined codebook. However, the practical \ac{ris} element coefficients might differ from the predefined values, due to electronic component characteristics drift over temperature and age. Furthermore, \ac{mc} and uncertainty
of the phase shifter control network would also
introduce inhomogeneities among \ac{ris} RF branches~\cite{wang2021over}. These effects can influence both the amplitude and phase of RIS coefficients. To capture this effect, we model the actual \ac{ris} coefficient matrix, i.e., \ac{nc} matrix using a perturbed representation
\begin{equation}\label{eq:b_2}
\bv_{\text{NC}}(\phi) = \Tilde{\Wm}^\top\av(\phi) ,
\end{equation}
where  
\begin{equation*}
    \Tilde{\Wm} = \Wm \odot \Delta \Wm, \text{subject to } \left\| \tilde{\mathbf{w}}_g \right\| = 1
\end{equation*}
and \( \Delta \Wm \in \CC ^ {N \times G} \) represents the perturbation matrix introduced by practical imperfections. The non-ideal coefficient matrix $\Tilde{\Wm}$ is defined as $[\Tilde{\Wm}]_{n,g} = \Tilde{\rho}_{n,g}e^{j\Tilde{\beta}_{n,g}}$, where $\Tilde{\rho}_{n,g}$ and $\Tilde{\beta}_{n,g}$ represent the practical amplitude and phase adjustments at the $n$-th element for the $g$-th codeword, respectively. The constraint here ensures consistent reflected power across different codewords. 
\subsection{Beam Model with Combined Impacts}
The most comprehensive model in our study accounts for the \ac{ci} of element pattern, RF impairments, measurement uncertainties. Incorporating these effects, the beam representation is given by  
\begin{equation}\label{eq:b_all}
    \bv_{\text{CI}}(\phi) = \gamma(\phi) \Tilde{\Wm}^\top \av(\phi),
\end{equation}  
where NC matrix $\Tilde{\Wm}$ defined in \eqref{eq:b_2} accounts for the RF impairments and \ac{mc} effects. The correction coefficient $\gamma(\phi)$ represents the combination effects of the element pattern and the unbalanced channel gain due to the measurement uncertainties in an anechoic chamber. This formulation extends the ideal beam representation by integrating practical non-idealities, providing a more realistic characterization of the beamforming response. 

The unknown parameters in these beam models will be extracted using the proposed calibration algorithms, with the assistance of measured data, as described in Section \ref{sec:algorithm}. The calibrated beam patterns and the resulting localization performance based on different beam models will be analyzed in Section \ref{sec:validation}.  

\section{Beam Calibration}\label{sec:algorithm}
In this section, we first define a pattern similarity evaluation metric as an objective function to design calibration algorithms. Then, a mismatch analysis is derived to evaluate the impact of the calibration error on the localization performance.   
\subsection{Pattern Similarity Evaluation Metric}
An intuitive baseline here is that the calibrated beam should be as close as possible to the ground truth (i.e., the measured beam pattern in the anechoic chamber). Therefore, the evaluation metric which describes the similarity between the true beam and the calibrated beam can be defined as
\begin{equation}
    L_0 = \int_{-\pi/2}^{\pi/2} \Vert \bar\bv(\phi) - \bv(\phi) \Vert d\phi,
\end{equation}
where $\bar\bv(\phi)$ denotes the ground truth of the beam patterns obtained from the measurements shown in Section~\ref{sec:validation}-A. For simplicity, a discretized version of the calibration error can be formulated as
\begin{equation}\label{eq:loss_function}
    L_1 = \frac{\sum^S_{s=1}\sigma_{s}\Vert \bar\bv(\phi_s) - \bv(\phi_s)\Vert_2^2}{\sum^S_{s=1}\sigma_{s}},
\end{equation}
where a set of sample angles $\thetav_s = [\theta_1, \ldots, \theta_S]^\top$ are chosen to evaluate the calibrated beam. Depending on the area of interest, $\sigma_s$ can be delicately designed (e.g., large weights for boresight directions).
This metric is used as an objective function for the calibration algorithms in Section \ref{sec:algorithm}.
\subsection{Calibration Algorithms}
In practical implementations, the ground truth \ac{ris} beam patterns are sampled over a set of discrete angles through measurements in the anechoic chamber, denoted by $\bar{\Bm}(\phiv) \in \mathbb{C}^{T \times G}$ ($\phiv = [\phi_1, \phi_2, \ldots, \phi_T]^\top$). Consequently, the beam response vector $\bv(\phi)$ in the beam models presented is replaced by the beam response matrix $\Bm(\phiv)= [\bv(\phi_1), \bv(\phi_2), \ldots, \bv(\phi_T)]^\top \in \mathbb{C}^{T \times G}$. Similarly, the steering vector $\av(\phi)$ and the correction coefficient vector $\gamma(\phi)$ are substituted with the steering matrix $\Am(\phiv) \in \mathbb{C}^{N \times T}$ and correction coefficient matrix $\Gammam(\phiv) = \text{diag}(\gamma(\phi_1),\gamma(\phi_2), \ldots, \gamma(\phi_T))$, respectively. Note that the unknown parameter $\Tilde{\Wm}$ for the beam model with NC in \eqref{eq:b_2} can be directly estimated using the least squares method. Therefore, in the following discussion, we focus on the calibration algorithms for the beam models in \eqref{eq:b_1} and \eqref{eq:b_all}.
\subsubsection{Calibration Algorithm for Beam Model with MCM}
The updated beam model for the $g$-th codeword with \ac{mcm} is denoted as
\begin{align}
    \bv_\text{MCM}^{g}(\phiv) &= \Am^\top(\phiv) \Cm_{g} \wv_{g} 
    =\Mm_g \wv_g
    \label{eq:model_1_2},
\end{align}
where $ \bv_\text{MCM}^{g}$ is the beam response vector for the $g$-th codeword and $\Cm_g = \mathrm{Toeplitz}([1, c_{g,1}, c_{g,2}, 0, \ldots, 0])$. Higher-order coupling terms are neglected here. The matrix $\Mm_g$ can be estimated via least square, shown as
\begin{equation}
    \hat{\Mm}_g = \argmin_{\Mm_g} \| \bar\bv_g (\phiv) - \Mm_g \wv_g \|_2^2,
\end{equation}
where $\bar\bv_g (\phiv)$ is the ground truth beam with the $g$-th codeword. 
After we obtain $\hat{\Mm}_g$, we estimate the \ac{mcm} $\Cm_{g}$ based on $\hat{\Mm}_g = \Am^\top(\phiv) \Cm_{g}$. To this end, we first vectorize $\hat{\Mm}_g$ shown as
\begin{equation}
\label{m_g_Dc}
    \hat{\mv}_g \triangleq \Vec{(\hat{\Mm}_g)} = \Vec{(\Am^\top(\phiv) \Cm_{g})} = {\Dm}{\cv}_g,
\end{equation}
where ${\cv}_g \triangleq [1, c_{g,1}, c_{g,2}]^\top$ and ${\Dm} \triangleq ({\Id_{N}} \otimes \Am^\top(\phiv)){\Jm}$ with selection matrix $\Jm \in \mathbb{R}^{N^2 \times 3}$ defined as
\begin{align}
    [\Jm]_{i,j} = \left\{ \!\!
    \begin{array}{ll}
        1, & [\Vec{(\Cm_{g})}]_{i} = [\cv_{g}]_{j} \\
        0, & \text{otherwise},
    \end{array}
    \right.
\end{align}
for $i = 1, 2, \cdots, N^2$ and $j = 1, 2, 3$. To proceed, we formulate \eqref{m_g_Dc} in a real-valued manner, as $\hat{\mv}_{g, \text{R}} = {\Dm}_{\text{R}}{\cv}_{g, \text{R}}$, where
\begin{align*}
    \hat{\mv}_{g, \text{R}} \! \triangleq \!\! \left[ \!\!\!
    \begin{array}{l}
         \text{Re}(\hat{\mv}_{g})  \\
         \text{Im}(\hat{\mv}_{g})
    \end{array}
    \!\!\! \right] \!\!,
    {\Dm}_{\text{R}} \! \triangleq \!\! \left[ \!\!\!
    \begin{array}{lc}
       \text{Re}(\Dm)  & \!\! -\text{Im}(\Dm) \\
       \text{Im}(\Dm)  & \!\! \text{Re}(\Dm)
    \end{array}
    \!\!\! \right] \!\! ,
    {\cv}_{g, \text{R}} \! \triangleq \!\! \left[ \!\!\!
    \begin{array}{l}
         \text{Re}({\cv}_{g})  \\
         \text{Im}({\cv}_{g})
    \end{array}
    \!\!\! \right] \!\!.
\end{align*}
Then, ${\cv}_{g, \text{R}}$ can be estimated as $\hat{\cv}_{g, \text{R}} = ({\Dm}_{\text{R}}^\top{\Dm}_{\text{R}})^{-1}{\Dm}_{\text{R}}\hat{\mv}_{g, \text{R}}$, and \ac{mc} parameters $c_{g,1}$ and $c_{g,2}$ is estimated as
\begin{align}
    \hat{c}_{g,1} = [\hat{\cv}_{g, \text{R}}]_{2} + \jmath [\hat{\cv}_{g, \text{R}}]_{5} \text{~and~} \hat{c}_{g,2} = [\hat{\cv}_{g, \text{R}}]_{3} + \jmath [\hat{\cv}_{g, \text{R}}]_{6}.
\end{align}

\subsubsection{Calibration Algorithm for Beam Model with CI}
The updated beam model with CI is given as
\begin{equation}
    \Bm_\text{CI}(\phiv) = \mathbf{\Gamma}(\phiv) \Am^\top(\phiv) \Tilde{\Wm}
    \label{eq:model_3}.
\end{equation}
We aim to estimate $\Tilde {\Wm}$, and $\Gammam$ by minimizing the Frobenius norm of the residual
\begin{align}
 [\hat\Wm, \hat\Gammam] 
&= \argmin_{\Tilde{\Wm}, \Gammam} \| \bar\Bm(\phiv) - \mathbf{\Gamma}(\phiv) \Am^\top(\phiv) \Tilde{\Wm} \|_F^2 \notag \\
&\quad \text{subject to } \left\| \tilde{\mathbf{w}}_g \right\| = 1
\end{align}
It is not feasible to estimate the two unknown parameters without any prior information. Therefore, initial estimates are necessary to enable the alternating optimization procedure. For optimizing $\Tilde{\Wm}$ in the first stage, the ideal codebook matrix $\Wm$ could be a reasonable initial point. Using the loss function defined in \eqref{eq:loss_function}, this problem can be solved using an iterative optimization method. 
\subsubsection*{Stage 1: Update of the NC \(\Tilde{\Wm}\)}

In the first stage, we update the estimated NC \(\Tilde{\Wm}\). Given an estimate of \(\check{\Gammam}\), the calibration problem reduces to the following minimization:

\begin{equation}
\check{\Wm} = \argmin_{\Tilde{\Wm}} \left\| \bar\Bm(\phiv) - \check\Gammam(\phiv) \Am^\top(\phiv) \Tilde{\Wm} \right\|_F^2.
\end{equation}
Since each column of \(\Tilde{\Wm}\) is constrained to have a unit norm, the optimization can be performed independently for each column vector \(\Tilde{\wv}_g\) as
\begin{equation}
\min_{\Tilde{\wv}_g} \left\| \bar\bv_g(\phiv) - \check\Gammam \Am^\top(\phiv) \Tilde{\wv}_g \right\|^2 
\quad \text{subject to} \quad \| \Tilde{\wv}_g \| = 1.
\end{equation}
The closed-form solution to the above problem is given by

\begin{equation}
\check{\wv}_g = 
\frac{(\Am(\phiv) \Am^\top(\phiv))^{-1} \Am(\phiv) \check\Gammam^{-1} \bar\bv_g}
{\left\| (\Am(\phiv) \Am^\top(\phiv))^{-1} \Am(\phiv) \check\Gammam^{-1} \bar\bv_g \right\|}.
\end{equation}

\subsubsection*{Stage 2: Update of the Calibration Matrix \(\Gammam\)}

In the second stage, we update the estimate of the diagonal calibration matrix \(\Gammam\). Given \(\check{\Wm}\), the optimization problem can be formulated as
\begin{equation}
\argmin_{\Gammam} \left\| \bar\Bm(\phiv) - \Gammam(\phiv) \Am^\top(\phiv) \check{\Wm} \right\|_F^2.
\end{equation}
Since \(\Gammam\) is diagonal, each diagonal element \(\gamma_t\) can be estimated independently by solving equations shown as
\begin{equation}
\check\gamma_t = \argmin_{\gamma} \left\| \bar\Bm(\phi_t) - \gamma \Am^\top(\phi_t) \check{\Wm} \right\|^2_2.
\end{equation}
The iteration stops upon reaching the maximum count or meeting convergence criteria.
\subsection{Mismatch Analysis}
The objective function used for beam calibration cannot directly reflect the impact of beam pattern errors on localization. To quantify the impact of the mismatch error, a theoretical lower bound analysis can be employed. Given the estimation problem in Section \ref{sec:problem_statement}, we define the channel parameter vector and the state vector as $\boldsymbol{\eta} = [\alpha_l, \tau_l, \alpha_r, \tau_r, \phi_A]^\top$, and $\boldsymbol{\sv} = [x_u,y_u]^\top$, respectively. Note that $\theta_A $ and $z_u $ are not considered in the above parameters, since the 2D beam patterns are adopted and the \ac{isac} system is limited to a 2D plane. We denote $\mathbf{y} \in \mathbb{C}^{GK} = \rm{vec(\mathbf{Y})}$ 
as the received signal vector. The probability density function (PDF) of the true model can be expressed as $f_{\text{TM}}( \yv|\bar{\etav})$, where $\bar{\boldsymbol{\eta}}$ is the vector of the true channel parameters. Similarly, the PDF of the misspecified model using the calibrated beam response is expressed as $f_{\text{MM}}( \yv|\etav)$. Since $f_{\text{TM}}( \yv|\bar{\etav})\neq f_{\text{MM}}( \yv|\etav)$, the corresponding mismatched lower bound (MLB) can be expressed as \cite{fortunati2017performance}
\begin{align}
    \text{MLB}(\bar {\boldsymbol\eta}, {\boldsymbol\eta}_0) 
    & = \underbrace{\Am_{{\boldsymbol\eta}_0}^{-1}\Bm_{{\boldsymbol\eta}_0}\Am_{{\boldsymbol\eta}_0}^{-1}}_{=\text{MCRB}({\boldsymbol\eta}_0)} + \underbrace{(\bar{\boldsymbol\eta} - {\boldsymbol\eta}_0)(\bar{\boldsymbol\eta} - {\boldsymbol\eta}_0)^\top}_{=\text{Bias}({\boldsymbol\eta}_0)}, 
    \label{eq_LB_channel_parameters}
\end{align}
where ${\boldsymbol\eta}_0$ is the pseudo-true parameter vector, and the derivation of $\Am_{{\boldsymbol\eta}_0}, \Bm_{{\boldsymbol\eta}_0}$ can be found in \cite{fortunati2017performance}. 

The pseudo-true parameter vector is defined as the point that minimizes the Kullback-Leibler divergence between $f_\text{TM}(\yv|\bar {\boldsymbol\eta})$ and $f_\text{MM}(\yv| {\boldsymbol\eta})$ as
\begin{align}
    {\boldsymbol\eta}_0 = \arg \min_{\boldsymbol\eta} D_\text{KL}(f_\text{TM}(\yv|\bar {\boldsymbol\eta})\Vert f_\text{MM}(\yv| {\boldsymbol\eta})).
\end{align}
Based on~\cite{ozturk2022ris} with simplification, the pseudo-true parameter can be obtained as
\begin{equation}
    {\boldsymbol\eta}_0 
    = \arg \min_{{\boldsymbol\eta}} \Vert \epsilonv(\etav) \Vert^2 
    = \arg \min_{{\boldsymbol\eta}} \Vert \bar\muv(\bar {\boldsymbol\eta}) - \muv({\boldsymbol\eta})\Vert^2  
    \label{eq_pseudotrue_final},
\end{equation}
where $\bar\muv(\bar {\boldsymbol\eta})$ and $\muv({\boldsymbol\eta})$ are the noise-free received signal vectors using the true beam response and the employed beam model, respectively.
A practical solution is to estimate ${\boldsymbol\eta}_0$  using gradient-based methods initialized with the true value $\bar {\boldsymbol\eta}$. The \ac{mcrb} part in~\eqref{eq_LB_channel_parameters} is dependent on the \ac{snr}. For analytical tractability, we focus on the biased term, which encapsulates the fundamental mismatch between the two models~\cite{chen2023modeling}. The pseudo-true parameter of the state vector $\sv_0$ and its corresponding bias term $\text{Bias}(\sv_0)$ can be calculated based on ${\boldsymbol\eta}_0$ and the geometry relationship of the localization system. We refer to $\text{Bias}(\sv_0)$ as the \ac{alb} for clarity in visualization in Section~\ref{sec:validation}.

\section{Performance Analysis}\label{sec:validation}
In this section, we employ the calibration algorithms introduced in Section~\ref{sec:algorithm} to evaluate the effectiveness of the three beam models in Section~\ref{sec:beam_model}.
\subsection{Ground Truth Data}
Ground truth data are essential for evaluating the effectiveness of the proposed \ac{ris} beam model. Here, the beam pattern of an \ac{ris} prototype consisting of \(16 \times 16 = 256\) elements was measured. Each \ac{ris} element supports 2-bit phase resolution for beamforming control. Additional details about the \ac{ris} prototype can be found in~\cite{li2023design}.
The coefficients assigned to the \ac{ris} elements were configured based on 11 distinct scanning angles, with an elevation angle fixed at \(0^\circ\) and azimuth angles ranging from \(-50^\circ\) to \(50^\circ\) in increments of \(10^\circ\). The resulting \ac{ris} beam patterns were measured across the azimuth plane, ranging from \(-90^\circ\) to \(90^\circ\) in \(1^\circ\) steps. However, the angle range used to calculate the performance metric is from \(-40^\circ\) to \(40^\circ\).
These measured patterns serve as ground truth for estimating the unknown parameters in the proposed beam models and are incorporated into numerical simulations to evaluate localization performance.

\subsection{Numerical Simulation Settings}
We consider a 2D RIS-aided localization system with one \ac{ue} and one \ac{ris}. The simulation parameters are summarized in Table \ref{tab:settings}. The ground truth data, i.e., the measured \ac{ris} patterns, are used in the simulation to generate the received signal matrix $\Ym$. Note that $G = 11 \times 32 = 352$ indicates that the \ac{ris} scans 11 distinct angles, as determined from the measured data, and repeats the scanning process 32 times within the coherence time. We calculate the \ac{alb} as the \ac{ue} moves within an area of $10 \times 10$ m$^2$. 
\begin{table}[h]
\centering
\caption{Default Parameter Settings}
\begin{tabular}{|l|l|}
\hline
\multicolumn{2}{|c|}{\textbf{Signal \& Frequency Parameters}} \\ \hline
Center frequency                  & 30 GHz      \\ \hline
Bandwidth                         & 200 MHz        \\ \hline
Number of subcarriers             & $K$ = 128 \\ \hline
Number of symbols                 & $G$ = 11$\times$32 = 352    \\ \hline
Noise figure                      & 10 dB    \\ \hline
\multicolumn{2}{|c|}{\textbf{Geometry Parameters}} \\ \hline
BS position               &  $[0, 0]^\top$   \\ \hline
UE position               &   $x_u \in [-5, 5]$ $y_u \in [-4, 6]$   \\ \hline
Dimensions of RIS       & $N = 1 \times 16$         \\ \hline
RIS position    & $\pv_{r} = [0,6]^\top$      \\ \hline
\end{tabular}
\label{tab:settings}
\end{table}
\subsection{Results Analysis}
\begin{figure}[h]
\centering
\centerline{\includegraphics[width=0.74\linewidth]{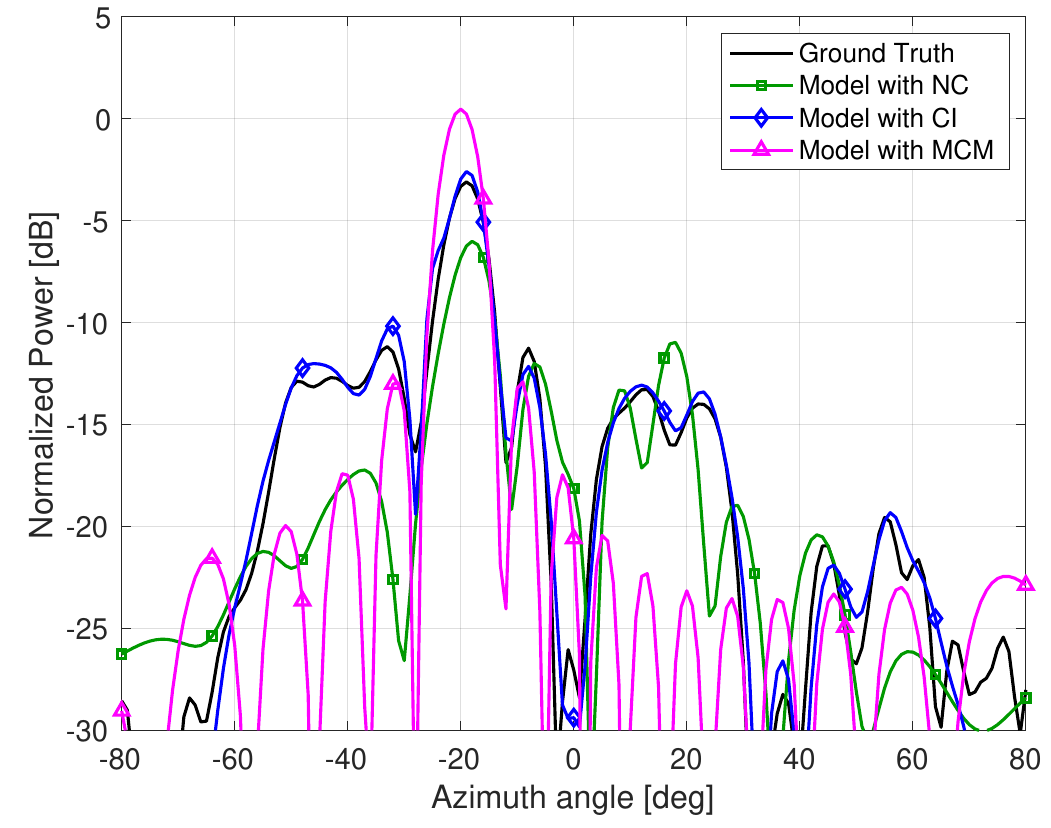}}
\vspace{-0.1cm}
\caption{Comparison of the measured RIS pattern and the calibrated pattern using different beam models (scanning angle at -20$^\circ$).}
\label{fig_pattern}
\vspace{-0.5cm}
\end{figure}

\begin{figure}[h]
\centering
\includegraphics[width=0.8\linewidth]{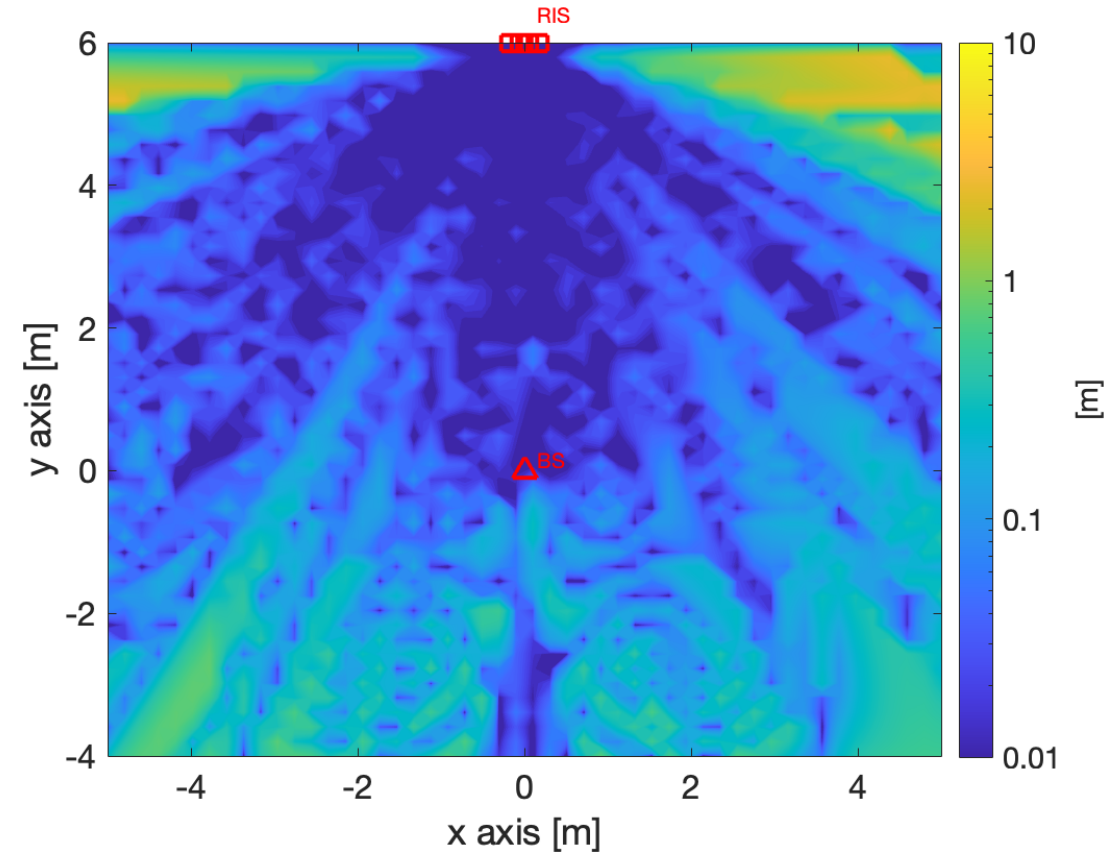}
\vspace{-0.1cm}
\caption{Visualization of \ac{alb} when \ac{ue} is located at different positions  within a 10 $\times$ 10 $\text{m}^2$ area using beam model with CI.}
\label{fig_heatmap}
\vspace{-0.2cm}
\end{figure}

\begin{figure}[h]
\centering
\includegraphics[width=0.76\linewidth]{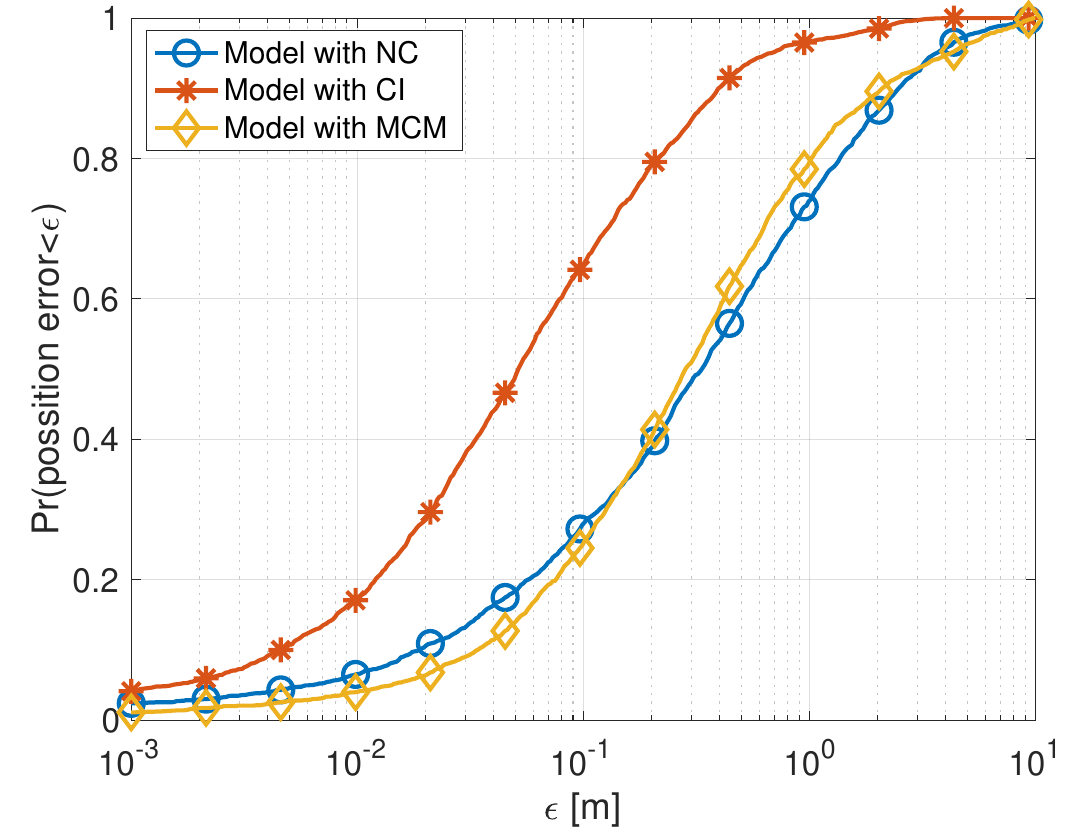}
\vspace{-0.2cm}
\caption{Comparison of the CDF for ALB within a 10 $\times$ 10 $\text{m}^2$ area using different beam models.}
\label{fig_CDF}
\vspace{-0.2cm}
\end{figure}

Using the calibration algorithms presented in Section~\ref{sec:algorithm}, the unknown parameters associated with each beam model can be accurately estimated. An example of the reconstructed $-20^\circ$ \ac{ris} beam patterns using different beam models is shown in Fig.~\ref{fig_pattern}, benchmarked by the ground truth pattern. 
The calibration errors, as defined in \eqref{eq:loss_function}, are 8.75, 4.15, and 0.94 for the beam models with \ac{mcm}, \ac{nc}, and \ac{ci}, respectively. It is evident that the beam model with CI outperforms the other two models in terms of beam similarity, demonstrating the effectiveness of both the model and the calibration algorithm. 
To directly investigate the effects of different calibrated beams on localization performance, we have calculated the \ac{alb} within a certain area using different calibrated \ac{ris} beams. 
The visualization of the \ac{alb} results using the beam model with CI is presented in Fig.~\ref{fig_heatmap}. It can be observed that \ac{alb} is large when the \ac{aoa} at \ac{ue} is large and \ac{ue} is far away from the \ac{ris} center. This is due to the path loss and the limited beamforming angle of the codebook. 
 Additionally, higher \ac{alb} values can be observed when the \ac{ue} is located in the gaps between the discrete scanning angles. This degradation arises from reduced signal power at these angular positions, leading to lower calibration accuracy of the beam pattern at these angles. The \ac{cdf} of the \ac{alb} results for different beam models are presented in Fig.~\ref{fig_CDF}. The beam model with \ac{ci} exhibits a high likelihood of achieving a positioning error below 1 m, demonstrating a notably greater improvement in localization accuracy compared to the other two models. In contrast, the beam models with \ac{mcm} and \ac{nc} generally achieve similar levels of localization accuracy. The CDF results further confirm the effectiveness of the beam model with CI.

\section{Conclusion}
\ac{ris} calibration is essential for achieving high localization accuracy in practical systems. In this work, we focus on calibrating the RIS beam to enhance localization performance. Three beam models are proposed, along with their corresponding calibration algorithms, to reconstruct the practical \ac{ris} beam pattern. The effectiveness of each beam model is evaluated through mismatch analysis in a typical RIS-aided localization scenario. The results demonstrate that the calibrated \ac{ris} beam using the CI-equipped beam model aligns closely with the measured ground truth, significantly outperforming the other two models in terms of localization accuracy.

\bibliographystyle{IEEEtran}
\bibliography{main}

\end{document}